\documentclass[12pt,tightenlines,eqsecnum,floats,showpacs,nofootinbib,amsmath,amssymb,aps,prd,superscriptaddress]{revtex4}

\usepackage{graphicx}
\usepackage{amsmath,amssymb}

\def\be{\begin{equation}}
\def\ee{\end{equation}}
\def\beq{\begin{eqnarray}}
\def\eeq{\end{eqnarray}}
\def\f{\frac}
\def\tf{\tfrac}

\def\d{{\rm d}}

\def\la{\langle}
\def\ra{\rangle}
\def\nn{\nonumber}

\begin{document}
%\preprint{\vbox{\baselineskip=12pt \rightline{IGC-11/03-X} }}

\title{Pre-Big-Bang Cosmology and\\ Circles in the Cosmic Microwave Background}

\author{William Nelson} \email{nelson@gravity.psu.edu}
\affiliation{Center for Fundamental Theory, Institute for
Gravitation and the Cosmos, Physics Department,
Pennsylvania State University, University Park PA 16802, USA}
\author{Edward Wilson-Ewing} \email{wilson-ewing@cpt.univ-mrs.fr}
\affiliation{Centre de Physique Th\'eorique de
Luminy\footnote{Unit\'e mixte de recherche (UMR 6207) du CNRS et
des Universit\'es de Provence (Aix-Marseille I), de la
M\'editerran\'ee (Aix-Marseille II) et du Sud (Toulon-Var);
laboratoire affili\'e \`a la FRUMAM (FR 2291).},
Case 907, F-13288 Marseille, EU}
\affiliation{Center for Fundamental Theory, Institute for
Gravitation and the Cosmos, Physics Department,
Pennsylvania State University, University Park PA 16802, USA}

\begin{abstract}

We examine the possibility that circles in the cosmic microwave background could be
formed by the interaction of a gravitational wave pulse emitted in some pre-big-bang
phase of the universe with the last scattering surface.  We derive the expected size
distribution of such circles, as well as their typical ring width and (for concentric
circles) angular separation.  We apply these results in particular to conformal cyclic
cosmology, ekpyrotic cosmology as well as loop quantum cosmology with and without
inflation in order to determine how the predicted geometric properties of these
circles would vary from one model to the other, and thus, if detected, could allow
us to differentiate between various pre-big-bang cosmological models.  We also obtain
a relation between the angular ring width and the angular radius of such circles
that can be used in order to determine whether or not circles observed in the cosmic
microwave background are due to energetic pre-big-bang events.

\end{abstract}


\maketitle

\section{Introduction}
\label{s1}

It has recently been claimed that the Cosmic Microwave Background (CMB) exhibits circles
of anomalously low temperature variance~\cite{Gurzadyan:2010da}. While the statistical
significance of this result is in doubt~\cite{we,msz,Hajian:2010cy} (see
however~\cite{Gurzadyan:2010xj,Gurzadyan:2011ac}),
the intriguing possibility that such circles could be signatures of a pre-big-bang phase
of our universe has been put forward~\cite{Gurzadyan:2010da}.  The idea, originally
introduced within the context of conformal cyclic cosmology~\cite{penrose}, is that
powerful events in the pre-big-bang epoch such as supermassive black hole collisions
could have produced pulses of gravitational waves which would then propagate to the
current post-big-bang epoch of our universe and intersect the observed CMB sphere where
these pulses would leave imprints appearing as circles to observers~\cite{Gurzadyan:2010da,penrose}.

More specifically, if a cosmological model allows a pre-big-bang epoch during which
the universe is large and classical, it could be populated by galaxies, stars and
black holes just as it is today and, in particular, it would be possible for
collisions between supermassive black holes to occur.  Such a collision would
release a brief, intense gravitational wave pulse that would expand radially at
the speed of light.  This pulse would travel through the transition between the 
pre- and post-big-bang phases and continue expanding in our current post-big-bang epoch and could
potentially be observed by future gravitational wave detectors.  In addition,
it has been suggested that this pulse could leave an imprint on the last scattering
surface (LSS).  This imprint could be detected as the temperature variance on the
surface of the pulse sphere would be slightly lower than average since the pulse
would homogenise the electromagnetic matter field and thus ensure that the temperature
is more highly correlated on the pulse sphere than elsewhere \cite{Gurzadyan:2010da, penrose}.
Finally, if this sphere with an anomalously low temperature variance intersects the observed
CMB sphere, it would appear to us as a circle in the CMB.

Even though such a scenario is rather speculative, the possibility that information from
a previous phase of cosmology can be gleaned from the CMB is an exciting one.  Note that
even if there are strong gravitational wave pulses coming from a pre-big-bang epoch that
interact with the CMB (which is already a strong assumption), it is not clear what
mechanism would allow these pulses to leave an imprint on the CMB. Indeed the standard Sachs-Wolfe
effect would simply change the average CMB temperature of the rings%
\footnote{Note that the results presented in this work concerning the geometric properties
of the circles are also applicable to rings that have an anomalously high or low average
temperature.  However, there is currently no observational data suggesting that such rings
exist.}
(rather than the variance) \cite{zibin} and hence some entirely new mechanism would be
required in order for the pulses to leave an imprint of type reported
in~\cite{Gurzadyan:2010da,penrose}.

Nonetheless, if such circles are indeed found, their properties would depend on the dynamics
of the pre-big-bang era of the universe and therefore different cosmological models would
predict different properties for circles in the CMB.  For example, if in one cosmology
black hole collisions are expected to occur close to the transition between the pre- and
post-big-bang epochs, the gravitational wave pulse sphere will not have much time to
expand and therefore would be smaller in this cosmology than in a different model where
the gravitational wave pulse has a longer time to expand before it reaches the last
scattering surface.  This is one example of a difference between various cosmological
models that will have an effect on the geometric properties of any circles in the CMB.
The goal of this paper is to determine some of the properties of these circles that
depend on the pre-big-bang cosmology, should conclusive observations be made.

The outline of the paper is as follows: in Sec.\ \ref{s2} we study the basic properties
of the gravitational wave sphere produced by a supermassive black hole collision and
also the CMB sphere seen today.  In Sec.\ \ref{s3}, we derive some relations between
geometric characteristics of the circles ---specifically their size, their width and
the separation between concentric circles--- and the properties of the spheres studied
in Sec.\ \ref{s2}.  Then in Sec.\ \ref{s4} we consider four specific cosmological
models which each have a pre-big-bang epoch: conformal cyclic cosmology, ekpyrotic
cosmology and loop quantum cosmology with and without inflation.  We use the results
of Sec.\ \ref{s3} in order to determine what circles in the CMB would appear like
for each model.  We end with a discussion in Sec.\ \ref{s5}.

\section{The Two Intersecting Spheres}
\label{s2}

\begin{figure}
\begin{center}
\includegraphics[scale=1.0]{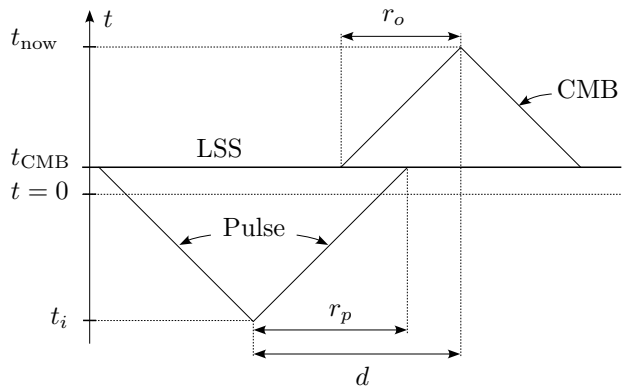}
\caption{\label{fig1} The conformal diagram of the expanding (null) pulse, intersecting
the last scattering surface (LSS).  The CMB sphere as seen today is also pictured.  The
time $t=0$ corresponds to the transition between the pre-big-bang and post-big-bang
epochs.  The detailed dynamics near the transition point are of course model-specific.}
\end{center}
\end{figure}

In order for there to be a circular imprint in the CMB, we first assume that an extremely
energetic event occurs in the pre-big-bang epoch and releases a strong pulse of
gravitational radiation.  This pulse forms an expanding sphere which propagates to the
current post-big-bang era and eventually reaches the LSS. To us the LSS appears to be
a sphere, with the Earth at the centre, simply because it occurred at the same time
everywhere due to homogeneity and the light travelled equal distances in each direction
due to the isotropic expansion of the universe.  Thus when the two spheres corresponding
to the gravitational wave pulse and the observed CMB intersect, the result is a circle.
See Fig.~\ref{fig1} for a conformal diagram showing the intersection of these two
spheres.

The interaction of the pulse with the recombination process taking place at the LSS
is then assumed to affect the plasma on these circles in some way which homogenises the
temperature, and this as yet unknown mechanism will then produce circles of abnormally
low temperature variance in the CMB which could potentially be observed \cite{penrose}.

We will now determine some of the relevant properties of the gravitational wave pulse
sphere and the observed CMB sphere.

\subsection{Some Properties of the Intersecting Spheres}
\label{s2.1}

Assuming homogeneity and isotropy as well as spatial flatness gives the flat
Friedmann-Lema\^itre-Robertson-Walker (FLRW) space-time whose line element
is given by
\be \label{metric} \d s^2 = - c^2 \d t^2 + a^2(t) \Big[ \d r^2 + r^2
\d \Omega^2 \Big], \ee
where $t$ is the proper time and we will normalise the scale factor so that
$a(t_{\rm now}) = 1$ where $t_{\rm now}$ is the proper time today.  Observations
indicate that recombination occurred approximately 380 000 years after the
beginning of the current era of the universe~\cite{weinberg}, or
\be \label{cmbt} t_{\rm CMB} \approx 10^{56}~t_{\rm Pl}, \ee
where we have set $t = 0$ to be the beginning of the current epoch of the
universe.  One can also show that the energy density at recombination is of
the order of
\be \label{cmbrho} \rho_{\rm CMB} := \rho(t_{\rm CMB}) \approx
10^{-115}~\rho_{\rm Pl}. \ee
Finally, since we have chosen $a( t_{\rm now} ) = 1$, we also find that
\be \label{cmba} a(t_{\rm CMB}) \approx 1089^{-1}. \ee
See, e.g., \cite{weinberg} for further background information about the
standard cosmological picture.

From~\eqref{metric}, it follows that the trajectory of a (null)
radial gravitational wave  pulse will be given by
\be c \f{\d t}{a(t)} = \d r. \ee
Therefore, the coordinate radius $r_p(t_f)$ of a gravitational wave sphere which
left a central point at $t_i$ (e.g., due to a supermassive black hole collision
occurring at $t_i$) is given by
\be\label{eq:coord_dist} r_p(t_f) = c \int_{t_i}^{t_f} \f{\d t}{a(t)}. \ee
Since the scale factor $a(t)$ is by definition positive, it follows that $r_p$
grows with time as one would expect.  Note that we shall use a lower case $r$ to
denote coordinate radii and an upper case $R$ for physical radii.  The physical
radius of the pulse sphere is simply given by
\be R_p(t_f) = a(t_f) \: r_p(t_f). \ee
As $a\left( t_{\rm now} \right) = 1$, it follows that today the physical and
coordinate distances coincide.

The coordinate and physical radii of a gravitational pulse due to an event which
occurred at the time $t_i$, when recombination occurs, is then given by
\be \label{eq:coord1} r_p : = r_p(t_{\rm CMB}) = c \int_{t_i}^{t_{\rm CMB}} \!
\f{\d t}{a(t)}, \qquad R_p(t_{\rm CMB}) = a\left( t_{\rm CMB} \right) r_p~. \ee
Similarly, the coordinate and physical radii of the CMB sphere as seen by an
observer today are
\be \label{eq:coord2} r_o := r_o(t_{\rm now}) = R_o(t_{\rm now}) = 
c \int_{t_{\rm CMB}}^{t_{\rm now}} \f{\d t}{a(t)}. \ee
It is possible to obtain a value for $r_o$ by using the Friedmann equation
\be\label{eq:H1}
 H^2 = \frac{8\pi G}{3} \left[ \frac{\rho_{\rm m}|_0}{a^3} + \frac{\rho_\gamma|_0}{a^4}
+\rho_\Lambda \right]~,
\ee
from which, using the $\Lambda {\rm CDM}$ parameters with $\rho_{\rm m}|_0\approx 2.12 \times
10^{-30}~{\rm g}~{\rm cm}^{-3}$ the (cold dark) matter density today, $\rho_\gamma|_0 \approx
7.8 \times 10^{-34}~{\rm g}~{\rm cm}^{-3}$ the radiation density today and the energy density
due to the nonzero cosmological constant being $\rho_\Lambda \approx 1.4 \times
10^{-29}~{\rm g}~{\rm cm}^{-3}$ (see, e.g., \cite{peter_uzan} for details), one finds that
\be\label{eq:H2}
 \dot{a}\approx 2\times 10^{-20}\left[ \frac{1}{a^2} + \frac{2700}{a} + 18000 a^2
 \right]^{1/2} {\rm s}^{-1}~.
\ee

Equation \eqref{eq:coord2}, rewritten in the form
\be
 r_o = c \int^{a(t_{\rm now})}_{a(t_{\rm CMB})} \! \frac{1}{a} \frac{{\rm d}t}{{\rm d} a}
 {\rm d} a~,
\ee
can be evaluated by using~(\ref{eq:H2}) and numerically integrating between
$a(t_{\rm CMB}) \approx 1089^{-1}$ and $a(t_{\rm now})=1$.  This gives the result
\be\label{ro}
 r_o \approx 3 \times 10^{61} \ell_{\rm Pl}~.
\ee

A final point is that recombination does not occur instantaneously but occurs
over a finite amount of time (between scale factors given by
$a = (1089 \pm 195)^{-1}$ \cite{peter_uzan}) and thus the last scattering surface
has a thickness of
\be \label{dro} \delta r_o \approx 3 \times 10^{59} \ell_{\rm Pl}. \ee

The value of $r_o$ and its thickness are both well known and will not change
from one pre-big-bang cosmological model to another.  However, $a(t_i)$ and
$r_p$ both strongly depend upon the cosmology.  This dependence of $a(t_i)$
and $r_p$ on the cosmological model will in turn have an effect on the
properties of any circles in the CMB.  This is what we shall quantify in
Sec.\ \ref{s3} before performing some case studies in Sec.\ \ref{s4}.
However, we will first examine some important geometric relations between
the intersecting gravitational pulse sphere and the CMB sphere with the
resulting circle.

\subsection{Geometric Relations Between the Spheres and the Resulting Circle}
\label{s2.2}

We will use the indices $p$ for the pulse sphere and $o$ for the observed
CMB sphere.  Finally, $\theta_p$ will denote the
angular radius of the circle as seen from the centre of the gravitational
wave pulse while $\theta_o$ will denote the circle's angular radius as
seen by us, the observers.  Note that these angles are defined as the
angular distance from a point on the circumference of the circle to
the centre of the circle.  Since the angles $\theta_i$ are a measure
of the radius of a circle rather than its diameter, it follows that
$\theta_i \in (0, \tf{\pi}{2})$.

It is immediately clear that
\be \label{sine} r_c = r_o \sin \theta_o = r_p \sin \theta_p~, \ee
and also that the coordinate separation $d$ between the centres of the
pulse sphere and the CMB sphere is given by
\be \label{secondA} d = r_p \cos \theta_p + r_o \cos \theta_o. \ee
Finally, by the cosine law it follows that
%Combining the two equations above via $\cos \theta_p = \pm \sqrt{1 - \sin^2 \theta_p}$,
%one finds
%%
%\be r_o \cos \theta_o - d = \mp \sqrt{r_p^2 - r_o^2 + r_o^2 \cos^2 \theta_o}, \ee
%%
%and then, by squaring both sides, it follows that
%
\be \cos \theta_o = \f{r_o^2 - r_p^2 + d^2}{2r_o d}. \ee
Note that if $r_p > r_o$, it is possible for $\cos \theta_o$ to be negative and
therefore for $\theta_o > \tf{\pi}{2}$.  However, the angular radius of such a
circle would in fact be given by $\pi - \theta_o$.  In order to incorporate this
effect, it is sufficient to change the above expression to
\be \label{cosine} \cos \theta_o = \f{|r_o^2 - r_p^2 + d^2|}{2r_o d}, \ee
and now $0 \le \theta_o \le \tf{\pi}{2}$ as desired.  Note that while this expression
holds for each circle, $r_p$ and $d$ will of course vary from circle to circle.

\section{Expected Characteristics of the Circles}
\label{s3}

\begin{figure}
\begin{center}
\includegraphics{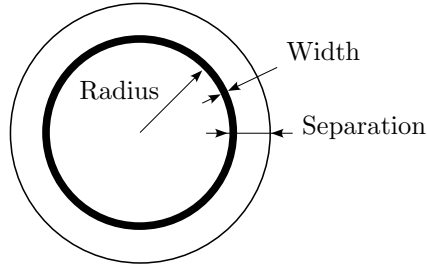}
\caption{\label{fig2} The geometric properties of the circles we are interested
in: their angular radius (Sec.\ \ref{s3.1}), the typical angular separation between
concentric circles (Sec.\ \ref{s3.2}) and the angular ring width of the circles
(Sec.\ \ref{s3.3}).}
\end{center}
\end{figure}

Here we will use the relations obtained in the previous section in order
to determine, for a given $a(t_i)$ and $r_p$, some of the characteristics we
expect the circles to have.  In particular, we will determine the expected
size of the circles (in terms of their angular radius), the typical angular
separation between concentric circles and also the angular ring width of the
circles (see Fig.~\ref{fig2}).  This will allow us to quantify how the
geometric properties of the circles vary when (the cosmological model and
hence) $a(t_i)$ and $r_p$ change.

\subsection{The Size of the Circles}
\label{s3.1}

Given $r_p$ and $d$, one can use~\eqref{cosine} to determine the angular
size of the circle.  While estimates for $r_p$ can be obtained for each
cosmology (see Sec.~\ref{s4} for some examples), $d$ is not an observable.
Since $d$ cannot be measured, it is impossible to use~\eqref{cosine}
to predict the size of a specific circle and hence what would actually be
observed in the CMB. However, it can be used in order to obtain the
{\it probability distribution} of the size of the circles, which could then
be compared to observations in order to constrain the pre-big-bang cosmological
model.

If we assume that the events that caused the gravitational wave pulses, e.g.,
supermassive black hole collisions, are spread out in an approximately random
fashion throughout the universe, then we can treat $d$ as an essentially
stochastic variable.  Then the expectation value for $d$ is given by
\be \la d \ra = \bigg[ \int_0^{4 \pi} \!\!\!\! \d \Omega \int_{d_{min}}
^{d_{max}} \!\!\!\!\!\!\!\!\! \d d \: d^2 \cdot d \bigg] \times \bigg[
\int_0^{4\pi} \!\!\!\! \d \Omega \int_{d_{min}}^{d_{max}} \!\!\!\!\!\!\!\!\!
\d d \: d^2\bigg]^{-1}, \ee
where $d_{min}$ and $d_{max}$ are easily determined as the gravitational
wave pulse sphere and the CMB sphere will only intersect if
\be\label{eq:inequ} |r_p - r_o| \le d \le r_p + r_o, \ee
and in the extreme cases the spheres only intersect at a point (see Fig.\
\ref{fig1}).

However, since $d$ cannot be measured, $\la d \ra$ is not a very useful quantity to
know.  Rather, we are interested in calculating the angular size of the circles in
the CMB.  Using~\eqref{cosine}, it can be seen that
\be \la \theta_o \ra = \bigg[ \int_0^{4 \pi} \!\!\!\! \d \Omega \int_{|r_p - r_o|}
^{r_p + r_o} \!\!\!\!\!\!\!\!\! \d d \: d^2 \cdot \theta_o(d) \bigg] \times \bigg[
\int_0^{4\pi} \!\!\!\! \d \Omega \int_{|r_p - r_o|}^{r_p + r_o} \!\!\!\!\!\!\!\!\!
\d d \: d^2\bigg]^{-1}, \ee
where
\be \theta_o(d) = \cos^{-1} \left( \f{|r_o^2 - r_p^2 + d^2|}{2 r_o d} \right), \ee
and now it is possible to determine the expected size of the circles in the CMB
if $r_p$ is known.

It is also straightforward to calculate the variance,
\be \sigma_{\theta_o}^2 = \la \theta_o^2 \ra - \la \theta_o \ra^2, \ee
where, as usual,
\be \la \theta_o^2 \ra = \bigg[ \int_0^{4 \pi} \!\!\!\! \d \Omega \int_{|r_p - r_o|}
^{r_p + r_o} \!\!\!\!\!\!\!\!\! \d d \: d^2 \cdot \theta_o^2 (d) \bigg] \times
\bigg[ \int_0^{4\pi} \!\!\!\! \d \Omega \int_{|r_p - r_o|}^{r_p + r_o} \!\!\!\!\!\!\!\!\!
\d d \: d^2\bigg]^{-1}. \ee
With these equations, we can predict the distribution of the size of the circles
in the CMB for a given cosmological model.

\subsubsection{The $r_p \gg r_o$ Limit}
\label{s3.1a}

Consider first the $r_p \gg r_o$ case, in which case $d \in (r_p - r_o, r_p + r_o)$ and
$d$ can be parametrised as
\be d = r_p + \ell r_o, \ee
where $\ell \in (-1, 1)$.

We can now evaluate the expected mean value of $\theta_o$,
\be \la \theta_o \ra = \f{ \int_{-1}^1 \d \ell \: r_o r_p^2 (1 + 2 \tf{r_o}{r_p} \ell
+ \tf{r_o^2}{r_p^2} \ell^2) \theta_o(\ell) }{2 r_p^2 r_o \left( 1 + \tf{r_o^2}{3 r_p^2}
\right)}, \ee
where
\be \theta_o(\ell) = \cos^{-1}\bigg[ \Big| \ell + \f{r_o}{2 r_p} \left( 1 -
\ell^2 \right) - \f{r_o^2}{2 r_p^2} \ell \left( 1 - \ell^2 \right) + O \left(
\f{r_o^3}{r_p^3} \right) \Big| \bigg], \label{cos-bigrp} \ee
and Taylor expansions have been used on the right hand side in order to obtain a
polynomial in $\tf{r_o}{r_p}$ inside the inverse cosine.  We can further simplify the
equation by expanding the inverse cosine itself around the point $\ell$.  However,
one cannot use the na\"ive Taylor expansion as it fails near $\ell \approx -r_o / 2 r_p$
when the argument of the inverse cosine is zero as the absolute value function is not
differentiable at zero.  In order to avoid this problem, we will use two separate
expansions: one for the region where $\ell$ is of the order of $r_o/r_p$ and another
elsewhere.

More precisely, for $\ell \in (-r_o / r_p, r_o / r_p)$, we find
\be \label{invCos1a} \theta_o(\ell) = \f{\pi}{2} - \left| \ell + \f{r_o}{2 r_p} \right|
+ O\left( \f{r_o^3}{r_p^3}\right), \ee
while for all other $\ell$
\be \label{invCos1b} \theta_o(\ell) = \cos^{-1} |\ell| - \f{r_o}{2 r_p} {\rm sgn}(\ell)
\sqrt{1 - \ell^2} + \f{3 r_o^2}{8 r_p^2} |\ell| \sqrt{1 - \ell^2} + O \left( \f{r_o^3}{r_p^3}
\right). \ee
Note that both expansions agree at the points $\ell = \pm r_o / r_p$, at least up to
order $r_o^3/r_p^3$.  These expressions are both easily integrated and one finds that
\be \la \theta_o \ra = 1 - \f{4 r_o^2}{9 r_p^2} + O \left( \tf{r_o^3}{r_p^3} \right). \ee
This shows that the average angular radius for circles in the limiting case $r_p \gg r_o$
will be 1~rad or $57^\circ$.  In order to determine the variance we must also calculate
$\la \theta_o^2 \ra$, once more using~\eqref{invCos1a} and \eqref{invCos1b}, and after a
tedious but straightforward calculation one finds
\be \la \theta_o^2 \ra = \pi - 2 -\frac{4}{27}\left( 3\pi-4\right)
\f{ r_o^2}{r_p^2} + O\left( \tf{r_o^3}{r_p^3} \right), \ee
and therefore the variance is
\be \sigma_{\theta_o}^2 = \pi - 3 - \frac{4}{27} \left( 3 \pi -10 \right)
\f{r_o^2}{r_p^2} + O \left( \tf{r_o^3}{r_p^3} \right). \ee
Therefore, for cosmological models where $r_p \gg r_o$, although the variance is relatively
high, it is clear that the average value of $\theta_o$ would be near $57^\circ$ and most
$\theta_o$ would be in the interval $(35^\circ, 79^\circ)$.

\subsubsection{The  $r_p = r_o$ Case}
\label{s3.1b}

One can also consider the case where $r_p = r_o$, in which case
\be \theta_o(d) = \cos^{-1} \left( \f{d}{2 r_o} \right)~, \ee
and then it follows that
\be \la \theta_o \ra = \f{2}{3} \approx 38^\circ~, \ee
while
\be \la \theta_o^2 \ra = \f{6 \pi - 14}{9} \approx (42^\circ)^2~, \ee
and this gives a variance of
\be \sigma_{\theta_o}^2 = \f{2 \pi - 6}{3} \approx (18^\circ)^2~. \ee

The expected variance is relatively large, but it is nonetheless clear that
typical circles in the CMB would be quite large (i.e., greater than $\approx 20^\circ$)
for the $r_p \approx r_o$ case.

\subsubsection{The $r_p \ll r_o$ Limit}
\label{s3.1c}

As the last limiting case, we will consider the situation where $r_p \ll r_o$.
We can again parametrise $d$, this time as
\be d = r_o + \ell r_p~, \ee
where $\ell \in (-1, 1)$.

In this situation
\be
 \theta_o = \cos^{-1} \left[ \f{ 1 +\ell\f{r_p}{r_o} - \tf{1}{2}\left(1-\ell^2\right)
\f{r_p^2}{r_o^2} }{ 1+ \ell\f{r_p}{r_o}}\right]
=\cos^{-1} \left[ 1 - \f{\tf{1}{2}\left(1-\ell^2\right)
\f{r_p^2}{r_o^2} }{ 1+ \ell\f{r_p}{r_o}}\right],
\ee
and using the expansion $\cos^{-1} (1 - x) \approx \sqrt{2 x} +  O(x^{3/2})$, we
can expand to get
\be
 \theta_o = \sqrt{ 1 - \ell^2} \: \f{r_p}{r_o} \left[ 1 - \f{\ell r_p}{2 r_o}
 + O \left( \f{r_p^2}{r_o^2}\right)\right]~.
\ee
With this expression for $\theta_o$, it is easy to calculate that
\be
 \langle \theta_o\rangle = \f{\pi r_p}{4 r_o} + O \left( \f{r_p^3}{r_o^3}\right)~,
\ee
and
\be \label{eq:s3.1c}
 \langle \theta_o^2 \rangle = \f{2 r_p^2}{3 r_o^2} + O \left( \f{r_p^4}{r_o^4}\right)~.
\ee
It then follows that the variance is
\be
 \sigma^2_{\theta_o} = \left( \f{2}{3} - \f{\pi^2}{16}\right) \f{r_p^2}{r_o^2}
 + O \left( \f{r_p^4}{r_o^4} \right)~.
\ee

A final important point is that the maximal angular radius $\theta_o^{\rm max}$ possible
in this case is
\be \theta_o^{\rm max} = \sin^{-1} \left( \f{r_p}{r_o} \right) = \f{r_p}{r_o}
+ O \left( \f{r_p^3}{r_o^3} \right); \ee
clearly, one of the principle signatures of the $r_p \ll r_o$ limit is the presence of
\emph{only} very small circles.

\subsubsection{Further Remarks}
\label{s3.1d}

After the study of these three limiting scenarios, it is clear that it is only for cosmological
models where $r_p$ is smaller than $r_o$ that one would \emph{a priori} expect to see more
small circles in the CMB than large ones.  In all other cases the majority of the circles
observed in the CMB should be quite large, with an angular radius approximately in the
range of $\theta_o \in (20^\circ, 90^\circ)$.

Nonetheless, there may be some selection effects which come into play.  For example, it
may be hard to see very large circles as they would necessarily cross the equatorial plane
in the CMB and therefore significant portions of these circles would be masked by the
foreground.  It is also possible that there are some other effects which make smaller
circles easier to detect (one such effect is presented in Sec.\ \ref{s3.3}).  Still,
unless these selection effects are very strong, the results found in the three limiting
cases studied here can be very useful, as we shall see in Sec. \ref{s4}.

\subsection{Separation Between Concentric Circles}
\label{s3.2}

\begin{figure}
\begin{center}
\includegraphics{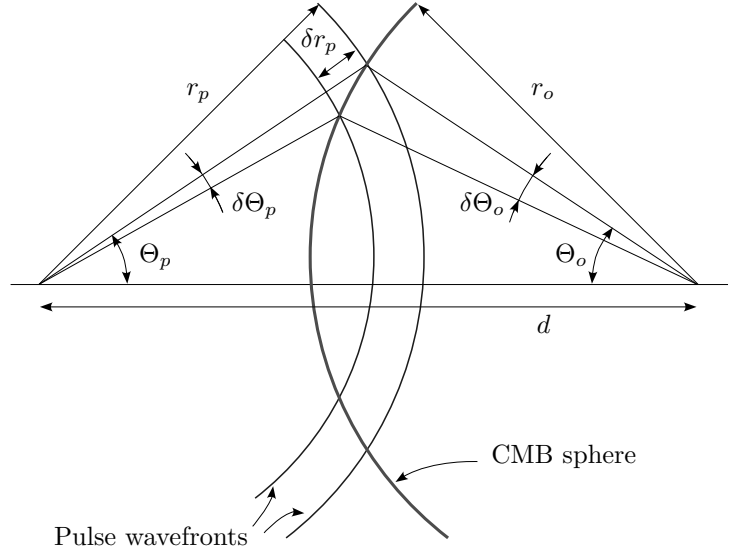}
\caption{\label{fig3} A pair of spherical pulses with coordinate radii $r_p$
and $r_p -\delta r_p$, intersecting the observed CMB sphere of radius $r_o$.}
\end{center}
\end{figure}

It has been pointed out that if there are repeated emissions from one particular
source (e.g., a supermassive black hole that collides with several other large black
holes over time), this would result in concentric circles \cite{Gurzadyan:2010da,
penrose}.  The angular separation between such concentric circles would depend
on the angular radius of the observed circles and the distance to the source as
well as the original frequency of the source.

For simplicity, we will assume recombination to have occurred instantaneously,
leading to a thin CMB surface (one can generalise the results presented here
by following similar arguments).  Let us consider two concentric pulse spheres,
of coordinate radii $r_p$ and $r_p - \delta r_p$, where $\delta r_p$ is the coordinate
distance between the two gravitational wave pulses.

Using the conventions shown in Fig.~\ref{fig3}, we can apply Eq.~\eqref{cosine}
to the larger and smaller concentric circles and this gives, respectively,
\beq
 \cos\theta_o &=& \frac{ |r_o^2 - r_p^2 + d^2|}{2r_o d}~, \label{sep1} \\
 \cos \left(\theta_o - \delta \theta_o\right) &=& \frac{ |r_o^2 -
 \left( r_p-\delta r_p\right)^2 + d^2|}{2r_o d}~. \label{sep2}
\eeq
Combining these equations gives
\be \label{eq:W2}
 \frac{\delta r_p \left( 2 r_p - \delta r_p \right)}{2r_o d} = \cos \left( \theta_o
 - \delta \theta_o \right) - \cos \left( \theta_o \right),
\ee
where for the sake of simplicity we have assumed that $r_o^2 - r_p^2 + d^2 > 0$.
Note that since the time elapsed between two separate gravitational wave pulses
can be significant, we cannot assume that $\delta r_p$ and $\delta \theta_o$ are
small%
% begin footnote
\footnote{Although one cannot assume $\delta r_p$ or $\delta \theta_o$ to be
small in generic situations, there are some cases where~\eqref{eq:W2}
can be simplified.  For example, if certain cosmological scenarios decree
that $r_p \ll r_o$, it follows that $\delta r_p$, being smaller than $r_p$,
is also much smaller than $r_o$ and also that $d \approx r_o$.  In this
limit, $\delta \theta_o < \theta_o \ll 1$ and, by using several Taylor
expansions, one can show that~\eqref{eq:W2} becomes $\delta r_p \approx
(r_o^2 / r_p) \cdot \delta \theta_o \sin \theta_o$.}.
% end footnote
If $r_o^2 - r_p^2 + d^2 < 0$, then the resulting equation will be a little
more complicated as the absolute value signs cannot be removed.

Following \eqref{eq:coord_dist}, we also find that the coordinate separation
$\delta r_p$ between two separate pulses is given by
\be \delta r_p = c \int_{t_1}^{t_2} \f{\d t}{a(t)}, \ee
where $t_1$ and $t_2$ are the (proper) times when, respectively, the first and second 
pulses were emitted.

In general, if there are a large number of circles observed in the CMB, one can use
this data to try to fit \eqref{eq:W2} in order to determine typical values of
both $\delta r_p$ and $r_p$.  This could provide some information on the pre-big-bang
epoch, in particular regarding the frequency of supermassive black hole collisions
in that era.

\subsection{The Width of the Circles}
\label{s3.3}

It is also possible to obtain some information regarding the widths of the circles.
If the cause of the circles is a pulse of gravitational waves intersecting the
last scattering surface, then the pulse width would result in some characteristic
width of the observed circles.

It is possible to use the same setup as in the previous section, where now the
two  waves in Fig.\ \ref{fig3} correspond to the leading trailing wave-fronts of
the same gravitational wave pulse, rather than two separate pulses, and
$\delta$ corresponds to the coordinate thickness of the pulse while $\delta
\theta_o$ gives the width of the circle.

It is also necessary to take the width $\delta r_o$ of the CMB sphere into account.
This situation can be described by a diagram similar to that given in Fig.\ \ref{fig3}
except that the CMB sphere on the right also has a width: the outer radius is given
by $r_o$, while the inner radius is $r_o - \delta r_o$.

A careful analysis shows that, depending on the situation, the relevant intersections
are either the $r_o$ and $r_p - \delta r_p$ intersection and the $r_o - \delta r_o$ and
$\delta r_p$ intersection or else the $r_o$ and $r_p$ intersection and the $r_o -
\delta r_o$ and $r_p - \delta r_p$ intersection%
\footnote{We would like to thank Alexey Bobrick for pointing this out.}.
We will begin by considering the first case.

Using the cosine rule for the first two intersections described above, we find that
\be \cos(\theta_o) = \f{d^2 + (r_o - \delta r_o)^2 - r_p^2}{2 (r_o - \delta r_o) d}, \ee
\be \cos(\theta_o - \delta \theta_o) = \f{d^2 + r_o^2 - (r_p - \delta r_p)^2}
{2 r_o d}. \ee
Since we expect the width of the circles to be quite small (say less than $1^\circ$),
we can restrict our attention to $\delta \theta_o \ll \theta_o$ and $\delta \ll r_p$
(we already know that $\delta r_o / r_o \approx 0.01$).  Therefore, we can solve for
$\delta \theta_o$ by taking the difference between these two equations and only keeping
terms that are linear in $\delta r_o, \delta r_p$ and $\delta \theta_o$.  This gives
\be \delta \theta_o \sin \theta_o = \f{r_p}{r_o d} \, \delta r_p +
\f{r_p^2 + r_o^2 - d^2}{2 r_o^2 d} \, \delta r_o. \ee
It is easy to study the other case as well and it can be incorporated by changing
the relation above to%
\footnote{By considering these two cases, we obtain this relation which holds for
both positive and negative $r_o^2 - r_p^2 + d^2$ so long as we use conventions
where $0 < \theta_o - \delta \theta_o < \theta < \tf{\pi}{2}$.}
\be \label{width1} \delta \theta_o \sin \theta_o = \f{r_p}{r_o d} \, \delta r_p +
\f{|r_p^2 + r_o^2 - d^2|}{2 r_o^2 d} \, \delta r_o. \ee
Note that since $\theta_o > \theta_o - \delta > 0$, it follows that $\theta_o$
is strictly greater than zero and therefore $\sin \theta_o > 0$.  The width of
the circles seen in the CMB comes from two contributions: the width of the
gravitational pulse $\delta r_p$ and the width of the CMB sphere $\delta r_o$
given in \eqref{dro}.  Since $\delta r_o$ is already known, the remaining task
is to determine $\delta r_p$.

If the circles observed in the CMB were to come from an event which has a 
typical time-scale $\delta t$, then assuming that the scale factor is
approximately constant during this event%
\footnote{An effect that has been neglected in this analysis is the fact
that the gravitational wavefront will spread as different wavelengths will
travel at different speeds through various matter fields.  Although this
effect is completely negligible when the matter energy density is low,
it could become important as the Planck scale is approached and the
electromagnetic and then the quark-gluon plasmas are formed.  This
would contribute to a slightly larger $\delta r_p$.}
\be \delta r_p = c \int_{t_i}^{t_i + \delta t} \f{\d t}{a(t)} \approx
\f{c \, \delta t}{a(t_i)}, \ee
where we have made the approximation that the scale factor is approximately
constant in the interval $(t_i, t_i + \delta t)$.  This gives
\be \label{widthboth} \delta \theta_o = \f{r_p}{r_o d \sin \theta_o} \,
\f{c \delta t}{a(t_i)} + \f{|r_p^2 + r_o^2 - d^2|}{200 \, r_o d \sin \theta_o}, \ee
where we have used that $\delta r_o / r_o = 1/100$.

As a specific example, consider the collision of supermassive black holes.
In this case, the distance between the leading and trailing wave fronts should
be of the same order as the length of the ``chirp'' gravitational wave burst
that occurs at the end of the merger of two black holes.  The physical
time scale of the chirp in proper time coordinates is typically of the order
of a few days, i.e., $\delta t \sim 8.6 \times 10^4$ s (the time scale may
change depending upon the exact type of event in which case one would have
to modify the following relations accordingly).  Therefore,
\be \label{drp} \delta r_p \approx \f{2 \times 10^{48}~\ell_{\rm Pl}} {a(t_i)}, \ee
where $a(t_i)$ is the scale factor of the universe when the black holes
merge.  From \eqref{dro}, it follows that, unless $a(t_i) \le 10^{-11}$
or the coefficient of $\delta r_p$ is larger than that of $\delta r_o$ by
several orders of magnitude, the contribution due to $\delta r_p$ is
completely negligible compared to that due to $\delta r_o$ in which
case
\be \label{width} \delta \theta_o \approx \f{|r_p^2 + r_o^2 - d^2|}{2 r_o ^2 d
\, \sin \theta_o} \, \delta r_o \approx \f{|r_p^2 + r_o^2 - d^2|}{200 \, r_o d
\, \sin \theta_o}. \ee
One can solve for $d$ in terms of $r_o, r_p$ and $\theta_o$ by using \eqref{sine} 
and \eqref{secondA}, but that will not be necessary for our purposes
here.

In order for a circle to be observed by the WMAP and/or Planck satellites, its width
must be at least as large as the angular resolution of the satellites which is of the
order of $0.25^\circ$.  Clearly, in some cosmological scenarios the width should
be observable, while in others it might not be.

We can also see that it is easier to detect smaller circles due to the denominator
of $\sin \theta_o$.  Indeed, the denominator of $\sin \theta_o$ provides a
significant selection effect whereby smaller circles will be thicker and therefore
easier to detect.  Therefore, even if the typical circle size for a given
cosmological model is quite large, small circles could be the first ones observed
as their width is the largest.  Assuming circles are detected in the CMB with
sufficient resolution, it should be possible to detect this variation in the
width with respect to the angular size of the circles%
\footnote{Note that there is an additional dependence on $\theta_o$ in $d$ as
$d = r_o \cos \theta_o + \sqrt{ r_p^2 - r_o^2 \sin^2 \theta_o}$.  However, so
long as $r_o$ and $r_p$ are of different orders of magnitude, this additional
dependence on $\theta_o$ will be sub-leading.}.

As a final remark, we point out that if $\delta \theta_o \ge 5^\circ$, the
approximation that lead to (\ref{width1}) breaks down.  In this case, one would
have to keep all of the higher order terms that were neglected in this analysis
in order to determine the relation between the thickness of the pulse and the
width of the circle.

\section{Examples: Specific Pre-Big-Bang Cosmological Models}
\label{s4}

Now we will consider four specific pre-big-bang cosmological models in order to determine
how the properties of circles appearing in the CMB vary from one model to another.
Specifically, we will derive typical values of $a(t_i)$ and $r_p$ for each cosmology
and determine in turn what circle size and width are to be expected in each case.

We will begin by considering conformal cyclic cosmology (CCC) where it was first pointed
out that gravitational waves from a pre-big-bang eon might leave circular imprints in
the CMB.  We will then go on to consider three other cosmological models which also
have a pre-big-bang branch and hence could also provide a similar mechanism that would
lead to circles in the CMB.  These models are the ekpyrotic universe as well as loop
quantum cosmology (LQC), both with and without slow-roll inflation after the bounce.

It is important to point out that much more work is needed in order to say that any
of these cosmological models necessarily predicts the presence of circles in the CMB.
In particular, one would need to understand how the perturbations in the gravitational
field will be transferred to the CMB and how strong this effect will be.  It is quite
possible that this effect will be very small and that therefore it will be impossible
to detect, even if it is indeed present. One would also need to have a fully developed,
non-singular theory in order to describe how these perturbations propagated from the
pre- to post-big-bang epochs. For example, although general symmetry arguments may be
sufficient to predict that the pulses remain spherical through the transition 
(at least assuming cosmological anisotropies remain small), only a full understanding of the
underlying theory would allows us to predict that such transitions even occur.
Thus, the presence of circles in the CMB is a signature
of a pre-big-bang cosmological scenario rather than a prediction of them.

Nonetheless, if such circles are observed, their properties would provide information
about the pre-big-bang phase, in particular regarding the typical values of $r_p$,
which is related to the average size of the circles, and $a(t_i)$, which can affect
the width of the circles.  This would, in principle, allow us to differentiate between
different pre-big-bang cosmological models.

\subsection{Conformal Cyclic Cosmology}
\label{s4.1}

Conformal cyclic cosmology  (CCC) was first introduced by Penrose and the basic idea is
that the universe, when the scale factor diverges to infinity, is conformally mapped to
a space-time where the scale factor is zero and hence to another big bang at which point
a new epoch begins \cite{penrose}.  Due to the presence of a nonzero cosmological
constant, the scale factor of the universe diverges in a finite amount of conformal
time and therefore there are an infinite number of epochs: each new epoch begins
with a big bang when the infinite scale factor of the previous epoch is conformally
mapped to zero.

There is an ambiguity in how this mapping is done which is still not fully understood
(see, e.g., Sec.\ B10 in \cite{penrose}) and here we will choose a conformal mapping
where the coordinate distance (and therefore also the comoving distance) between
any two points remains constant during the transition from one epoch to the next.
This is a conformal mapping with nice properties, but it is one possibility out
of a one-parameter family of allowed conformal mappings.  For this section, we will
focus on the mapping chosen here and we will comment at the end on other possible
choices.

In the CCC paradigm, the most powerful supermassive black hole collisions would
happen at very late times in each epoch and thus $a(t_i) > a(t_{\rm now}) = 1$.
Such a scale factor at the black hole collision time indicates that the width of
the circles is dominated by the thickness of the CMB sphere [see \eqref{dro},
\eqref{widthboth} and \eqref{drp}] and therefore Eq.~\eqref{width} holds and can
be used to relate the width of a given circle to $r_o, r_p$ and its angular radius.

It is possible to bound $r_p$,
\be r_p = \int_{t_i}^{t_{\rm CMB}} \f{\d t}{a(t)}, \ee
by determining the behaviour of $a(t)$ in each epoch.  At first, when the pulse
occurs, the universe's dynamics are dominated by the cosmological constant $\Lambda
\approx 3.3 \times 10^{-122} \ell_{\rm Pl}^{-2}$  \cite{peter_uzan} and therefore,
for $t \in (t_i, +\infty)$,
\be a(t) = \exp ( H t ), \ee
where the Hubble rate is $H = \sqrt{\Lambda/3}$.

For the next epoch, there is no inflation after the big bang and we will assume that
the universe is radiation-dominated, at least until recombination, in which case we
have the relation
\be \rho = \f{\rho_o}{a(t)^4}, \ee
where $\rho_o$ is determined by some initial conditions.  From \eqref{cmbrho} and
\eqref{cmba}, one can see that
\be \rho_o \approx 10^{-127} \rho_{\rm Pl}, \ee
and then by solving Einstein's equations, one finds that for $t \in (0, t_{\rm CMB})$
\be a(t) = \left( \f{32 \pi G \rho_o}{3} \right)^{1/4} \sqrt{t}. \ee
It follows that
\begin{align} r_p &= \int_{t_i}^{+\infty} \f{\d t}{e^{Ht}} + \left( \f{3}{32 \pi G \rho_o}
\right)^{1/4} \: \int_0^{t_{\rm CMB}} \f{\d t}{\sqrt{t}} \nn \\ & = \f{1}{a(t_i) H} +
\left( \f{3} {2 \pi G \rho_o} \right)^{1/4} \: \sqrt{t_{\rm CMB}}~. \end{align}
Since $a(t_i) > 1$, the above equation provides an upper bound to $r_p$ when we set $a(t_i) = 1$.
Using \eqref{cmbt}, it follows that
\be \label{rpccc} r_p \le 7.8 \times 10^{60}~\ell_{\rm Pl} + 5 \times 10^{59}~\ell_{\rm Pl}
\approx 8 \times 10^{60}~\ell_{\rm Pl}. \ee
This is smaller than $r_o$ by a factor of 3.75 [see \eqref{ro}] and therefore the results
presented in Sec.\ \ref{s3.1c} describe the expected distribution of the circle sizes up to
errors of a few percent.  In particular, the largest circle possible is one where
$\sin^{-1} \theta_o = r_p / r_o$ and this corresponds to%
\footnote{Note that in \cite{penrose} the maximum angular radius allowed is given as
$30^\circ$ rather than the $15^\circ$ we find.}
\be \theta_o^{\rm max} \approx 15^\circ. \ee
In addition, since in the limiting case $r_o \approx 3.75 r_p$ (i.e.\ the largest possible $r_p$)
we have $d \approx r_o$, Eq.\ \eqref{width} simplifies to
\be \label{widthccc} \delta \theta_o \approx \f{1}{3000 \sin \theta_o}~. \ee
 It follows that in this case the width of the circles could be detected,
but only for small circles whose angular radius is less than approximately $3^\circ$.  This
will of course slightly vary depending on the exact relation between $d$ and $r_o$.

Therefore, in CCC with the choice of the conformal mapping chosen above, any circular imprints
in the CMB would have at most an angular radius of approximately $15^\circ$ and whose relation
between the width and radius is given by \eqref{widthccc}.

Alternatively, one could choose a different conformal mapping where the conformal factor differs
by a factor $\alpha$ in order to obtain, e.g., a different maximal radius for the circles.  In
this case the first term in \eqref{rpccc} will be multiplied by $\alpha$ and this will
change the size distribution of the circles as well as the maximal circle size possible.  It
will also change the relation between the width and the size of the circles, although 
\eqref{width} will still hold so long as $\alpha < 10^{11}$.  However, the conformal mapping
that we used in this section appears to us to be the most natural one and a different choice
may be difficult to motivate.

\subsection{The Ekpyrotic Universe}
\label{s4.2}

The ekpyrotic universe is a cyclic cosmological model motivated by string theory \cite{kost, st1, st2}.
The idea is that the classical big bang corresponds to a collision between two branes and that
the accelerated expansion of the universe at late times is due to the potential energy between
the two branes being slightly larger than zero when the separation is large.  This period of
accelerated expansion lasts for a very long time and ensures that the entropy density and black
hole number density both become negligible before the big-crunch/big-bang transition and thus
set appropriate ``initial conditions'' for the beginning of the next cosmological epoch.  The
field that determines the distance between the two branes is modelled by a scalar field with
a potential that satisfies certain requirements, see \cite{kost, st1, st2} for further details.

Since the black hole number density in the universe at late times is very low, any supermassive
black hole collisions must happen around $\rho = \rho(t_{\rm now})$ in the previous epoch
and thus $a(t_i) \approx 1$.  It follows that, as in CCC, the observed width of any circles in
the CMB would be due to the width of the CMB sphere and Eq.~\eqref{width} holds.

In addition, given a specific potential for the scalar field, it is also possible to derive an
approximate value for $r_p$.  A portion of the calculation is very similar to the one performed
in the previous section on CCC: first there is a very long phase of accelerated expansion which
can be modelled by a small cosmological constant and then, after the big-crunch/big-bang
transition, there is a radiation-dominated phase up until the point where the last scattering
surface forms.  Therefore, the two terms that contribute to the CCC model also contribute for
the ekpyrotic universe.

However, we must also include the contracting phase of the ekpyrotic cosmology in our calculation
where the universe collapses from a scale factor which is very large to one that is zero on a
time-scale of $\Lambda^{-1/2}$, where $\Lambda \approx 10^{-122} \ell_{\rm Pl}^{-2}$ is the
observed value of the cosmological constant.  The main part of this collapse is potential-dominated
and we can approximate the scale factor by
\be a(t) = a_{\rm max}^{1 - \sqrt\Lambda (t-t_{\rm con})}, \ee
where $t_{\rm con}$ is the time at which the universe starts to contract and $a_{\rm max}$ is
the maximal scale factor reached.  It is easy to see that this contracting phase contributes
to $r_p$ a term given by%
\footnote{In our calculations we have neglected the portion of the evolution very close to the
transition point $a(t=0) = 0$ which is dominated by the kinetic energy of the scalar field.
However, this part of the universe's evolution is incredibly short and does not significantly
contribute to the value of $r_p$.}
\be \int_0^{\Lambda^{-1/2}} \!\!\!\!\! \f{\d t}{a_{\rm max}^{1 - \sqrt\Lambda t}} =
\f{1}{\sqrt\Lambda \ln a_{\rm max}}, \ee
which is negligible compared to terms of the order of $10^{60} \ell_{\rm Pl}$ since we
expect $a_{\rm max}$ to be several orders of magnitude greater than 1.  Therefore, this
phase of the evolution of the universe contributes a term which is negligible when
compared to the two other terms that were present in CCC [see \eqref{rpccc}] and thus
\be r_p \le 8 \times 10^{60}~\ell_{\rm Pl}. \ee

This is the same result as in the CCC model and the observational predictions of the models are
therefore identical: the largest circle should have a radius of approximately $15^\circ$ and the
relation between the size and the width of the circles should be observable
(assuming the circles are indeed seen) and have the form of, e.g.,
\be \delta \theta_o \approx \f{1}{3000 \sin \theta_o}, \ee
for the limiting case of $r_p = \tf{4}{15} r_o$.

\subsection{Loop Quantum Cosmology without Inflation}
\label{s4.3}

Loop quantum cosmology (LQC) is the result of applying the methods of loop quantum gravity to simple
space-times such as the homogeneous and isotropic FLRW models \cite{aps, mb, aa}.  For our purposes,
the most important feature is that in LQC the universe undergoes a quantum bounce when the scalar
curvature approaches the Planck scale. The pre- and post-bounce phases are extremely well
approximated by classical contracting and expanding cosmologies respectively so long as the
scalar curvature remains far from the Planck regime.

In particular, the Friedmann equation is modified due to quantum geometry effects and, when the matter
field can be described as a perfect fluid, it becomes \cite{vt}
\be \left( \f{\dot{a}}{a} \right)^2 = \f{8 \pi G}{3} \rho \left( 1 - \f{\rho}{\rho_c} \right), \ee
where the dot denotes differentiation with respect to the proper time $t$ and $\rho_c \approx 0.41
\rho_{\rm Pl}$ is the critical matter energy density where the bounce occurs.  One can see that
quantum geometry effects provide a repulsive force when the matter energy density reaches the Planck
scale.  Also note that the dynamics are symmetric around the bounce point where $\rho = \rho_c$.

The matter fields behave in the same way as in standard cosmology, e.g., for a radiation fluid
\be \label{rholqcnoinf} \rho = \f{\rho_o}{a(t)^4}, \ee
where $\rho_o$ is a constant to be determined by the initial conditions.  For the initial conditions
given at recombination in \eqref{cmbrho} and \eqref{cmba},
\be \rho_o = 10^{-127}~\rho_{\rm Pl}. \ee

One can consider an LQC model which began in the distant past as a large, approximately homogeneous
universe that contracted, bounced and expanded to the universe we see today without undergoing a
period of slow-roll inflation. Whilst such a model would have several important difficulties that
would need to be addressed ---such as why the bounce was approximately homogeneous, why the
perturbations are almost scale invariant, etc.--- here we shall ignore these issues and simply
study the properties of the circles that would be produced in this model.

We will now assume that, since there is no inflation, the universe is radiation-dominated so long as
the energy density is greater than $\rho_{\rm CMB} = 10^{-115} \rho_{\rm Pl}$, i.e., the energy
density that recombination occurs at.  It is then possible to use \eqref{rholqcnoinf} in order
to solve the modified Friedmann equation and this gives
\be \label{lqcnoinfa} a(t) = \left( \f{32 \pi G \rho_o}{3} t^2 + \f{\rho_o}{\rho_c} \right)^{1/4}. \ee
It is clear that the scale factor is strictly greater than zero at all times ---there is no singularity---
and also that it is time symmetric around the bounce point at $t = 0$.

We will assume that any astrophysical event that could be a source for a circle seen in the CMB
would have to occur at relatively low energy densities and in particular at a time when the energy
density is less than $\rho_{\rm CMB}$.  We are choosing this cutoff as we are assuming that the
strength of any major collisions that occur at energies where there exists a photon, proton and
electron plasma would be strongly damped by this plasma%
\footnote{Although it might be possible for a supermassive black hole collision to release a lot of
energy even if the collision has been damped by the presence of a plasma, an investigation along these
lines is outside of the scope of this paper.  Nonetheless, if this were to happen, the circle
size would be very small as in this case $r_p \le 10^{60} \ell_{\rm Pl} \ll r_o$ and $\theta_o^{\rm max}
\le 1 / 30$ rad.}.

Since we are only considering collisions that occur when $\rho < \rho_{\rm CMB}$, it follows that
$a( t_i ) > 10^{-3}$ and therefore the width of the circles is dominated by the contribution from
$\delta r_o$ and we can use \eqref{width}.

It is also possible to calculate a minimal value for $r_p$ by assuming that the collision happens
just as the plasma is forming in the pre-bounce phase.  One would of course expect most
supermassive black hole collisions to occur much earlier than the time reverse of recombination
in the pre-big-bang epoch, but this limiting event will provide a lower bound for $r_p$:
\be r_p \ge c \int_{-t_{\rm CMB}}^{t_{\rm CMB}} \f{\d t}{a(t)}. \ee
Using \eqref{lqcnoinfa} to evaluate the expression above, one finds that
\be r_p \ge 10^{60}~\ell_{\rm Pl}, \ee
which is smaller than $r_o$ by a factor of 30.  However, unlike for CCC and the ekpyrotic universe,
this is a \emph{lower bound} and it is easy to obtain $r_p$'s which are significantly larger by
simply considering collisions that occur at times much earlier than when the plasma forms. Thus
within this model, it is possible to obtain both small (when $r_p \ll r_o$) and large (when $r_p
\geq r_o$) circles.  However, in all cases smaller circles should be thicker (assuming their width
is observed) due to the selection effect due to the $\sin \theta_o$ in the denominator in
Eq.~\eqref{widthboth}.

\subsection{Loop Quantum Cosmology with Inflation}
\label{s4.4}

If one introduces an inflaton field into an LQC model, the quantum bounce will be followed by a phase
of standard slow-roll inflation. In addition to resolving the classical singularity, LQC naturally
sets the initial conditions for slow-roll inflation~\cite{Ashtekar:2009mm,Ashtekar:2011rm}.  An
important point here is that since the inflaton is not a perfect fluid, $a(t)$ is not necessarily
symmetric around the bounce point: it is possible to have a long period of slow-roll inflation after
the bounce without a corresponding period of slow-roll deflation before the bounce and vice versa.
Of course, a symmetric evolution in which there is exactly the same amount of deflation and inflation
is also a solution, however the likelihood of this occurring is very low.

In the following we will assume that there are $N_{\rm asym}$ efoldings of asymmetry between the
slow-roll deflationary and inflationary phases, i.e., that $N_{\rm inf} - |N_{\rm def}| =
N_{\rm asym}$ where $N_{\rm inf}$ is the number of efoldings of slow-roll inflation after the
bounce and $N_{\rm def}$ is the number of efoldings of slow-roll deflation before the bounce.

We will also denote the number of efoldings of super-inflation/ deflation as $N_{\rm LQC}$, which
occurs immediately around the bounce point, is approximately symmetric and is typically 2 or 3.
Then the scale factor at the end of slow-roll inflation is approximately $a\left( t_{\rm end}\right)
\approx a\left(t = 0\right)e^{N_{\rm LQC} + N_{\rm inf}}$, whilst the scale factor at the time
slow-roll deflation began is $a\left( t_{\rm beg}\right) \approx a\left( t=0\right) e^{N_{LQC} +
|N_{\rm def}|}$. Thus, in this model the dynamics of the scale factor can be highly asymmetric
if $N_{\rm inf}$ and $N_{\rm def}$ are significantly different.

If we estimate the amount of expansion that the universe underwent after slow-roll inflation (i.e.,
during standard cosmology) to be approximately $3\times 10^{28}$ \cite{peter_uzan}, then we have
\be\label{eq:e1}
 a\left( t=0 \right) \approx e^{-65.6 - N_{\rm inf} - N_{\rm LQC}}~,
\ee
where recall $a\left( t_{\rm now} \right) = 1$.

Defining $a\left( t_i\right)$ to have occurred $x$-efoldings before the beginning of the slow-roll
deflation period, we have
\be\label{eq:e2}
 a\left( t_i\right) \approx e^{x + |N_{\rm def}| + N_{\rm LQC}} a\left(t=0\right)~.
\ee
In this model, it is possible for $\delta r_p$'s contribution to the width be significant
if $x < N_{\rm asym} + 40$.  Typically, we would expect $x$ to be at least 65.6 in order
for the universe to be sufficiently diluted so that any black hole collisions would not
be significantly damped and, for this to occur and for the inequality to hold, we must
have an extremely asymmetric solution.  In this case, the contribution of $\delta r_p$ will
no longer be negligible and it will be necessary to use \eqref{widthboth} in order to
describe the width of the circles.

It is possible to obtain a strong lower bound for $r_p$ by studying the contribution due to
the inflationary phase which is given by
\be c \int_{t_{\rm start}}^{t_{\rm inf}} \f{\d t}{a(t)} \approx \f{1}{a\left( t_{\rm start}
\right) H_{\rm inf}} \approx 10^{64} \ell_{\rm Pl}, \ee
where $a(t_{\rm start}) \approx 10^{-58}$ is the scale factor at the beginning of inflation
having assumed 70 efoldings and $H_{\rm inf} \approx 10^{-6} \ell_{\rm Pl}^{-1}$ is the Hubble
rate during inflation.

It then follows that, since this is just one of several contributions to $r_p$,
\be r_p \ge 10^{64} \ell_{\rm Pl}, \ee
which is clearly significantly larger than $r_o$.

Since $r_p \gg r_o$, it follows that the relation between the width and the radius of
the circles given in \eqref{widthboth} can be simplified.  First, since $r_p \gg r_o$
it follows that $\theta_p \ll 1$ and then by using the cosine rule,
\begin{align} r_p^2 + r_o^2 - d^2 &= 2 r_p r_o \cos (\pi - \theta_o - \theta_p) \\
&\approx 2 r_p r_o \cos (\pi - \theta_o). \end{align}
This allows us to express the width as
\be \delta \theta_o \approx \f{c \delta t}{a(t_i) r_o \, \sin \theta_o}
+ \f{\cot(\theta_o)}{100}, \ee
and therefore we would expect to be able to detect the width of small circles.  As for the
other models studied here, smaller circles are expected to be thicker.

An important point here is that the results obtained for LQC with inflation depend
solely on the presence of the inflationary phase in our current epoch as the properties
of the pre-big-bang epoch (apart from its existence) are washed out by inflation.
Therefore, the conclusions obtained in this section hold for all pre-big-bang cosmological
models that include inflation.  For these models, we should expect to see circles whose
width is greater for smaller circles and their average angular radius should be $57^\circ$
with a variance of $(22^\circ)^2$.

\section{Discussion}
\label{s5}

For many of the pre-big-bang cosmological models that are currently being studied, it is
still not entirely clear how inhomogeneities propagate from the pre-big-bang era to the
present epoch.  Nonetheless, if a pre-big-bang phase does indeed provide the seeds for
the inhomogeneities we see today in our universe, there exists the possibility of observing
some signatures of the pre-big-bang era in our universe and, in particular, in the CMB.

Of course, depending on the specific pre-big-bang cosmological model, there may exist
several signatures of the previous eon that one could find in the CMB.  However, one
particular potential signature common to several models (and which is relatively easy to
search for) is the presence of circles with abnormally low temperature variance in the
CMB.  Such circles would be due to gravitational waves produced by supermassive black
hole collisions (or other extremely energetic events) from the pre-big-bang era.  Even
though the mechanism which allows for the transfer of the correlations from the
gravitational waves to the last scattering surface is not clear, it is possible
to predict some of the geometric properties of these circles given a pre-big-bang
cosmological model.  It is worth noting that different types of circles, such as
abnormally high or low temperature circles due to the Sachs-Wolfe effect, could
potentially also exist in the CMB \cite{zibin} and the results presented here hold
for the geometric properties of these other types of circles as well.

An important point is that we assume that quantum gravity corrections to the size
of circles in the CMB will be negligible and therefore we only calculate the dominant
classical contributions.  Nonetheless, quantum gravity plays a central role as it
provides the bridge between the pre- and post-big-bang epochs: without it, there
would be no way for the gravitational wave pulse to travel from one epoch to
the other as there would necessarily be a singularity.  Of course, it would be
interesting to see how the pulse might be affected by quantum gravity effects near
the transition point, but this question cannot be answered until the full dynamics
of quantum cosmology are understood.

In this paper, we have obtained a probability distribution for the size of the circles,
studied the expected separation between concentric circles and predicted the width of
the circles.  Note that this analysis is not complete: for example, we make no claims
regarding the total number of circles that one would expect to see in the CMB.
Nonetheless, by studying four specific cosmological models which have a pre-big-bang
epoch ---conformal cyclic cosmology, ekpyrotic cosmology and loop quantum cosmology with
and without inflation--- we have shown that the probability distribution of the circle
size varies from one cosmological model to another and this could potentially be used in
order to differentiate between cosmological models based upon careful observations of
the CMB.  We must stress that we are not predicting the presence of these circles in
the CMB.  Rather, what we have done is to show how, assuming the circles are present,
their geometric properties would differ from one cosmological model to another.

In particular, in conformal cyclic cosmology and the ekpyrotic universe, one would
expect there to be only very small circles with an angular radius of at most $15^\circ$,
while in LQC without inflation one would expect larger circles as well whereas for LQC
with inflation (and also all other pre-big-bang models that have an inflationary era), one
should expect a distribution of large circles with an average angular radius of $57^\circ$.

Also, in all of these models one expects the width and the radius of the circles to
be related.  The relations presented in Sec.\ \ref{s4} describing the ring width for
each cosmological model are often given for a specific value of $r_p$, but they can
easily be generalized.  In fact, these relations offer a simple way to test whether
the circles found in \cite{Gurzadyan:2010da} are due to extremely energetic events
in a pre-big-bang epoch or not: if the smaller circles are also wider, this would
provide strong evidence in support of the viewpoint presented in \cite{Gurzadyan:2010da}.
If not, then it seems more likely that the circles are simply statistical flukes or,
perhaps, are due to a completely different mechanism. If the circles described in
\cite{Gurzadyan:2010da} are indeed imprints from pre-big-bang events, their geometric
properties will give significant insight into the dynamics of the pre-big-bang era of
our universe and also the nature of the quantum gravity effects that were undoubtedly
important in the early universe.

\acknowledgments
%\ack
We would like to thank Alexey Bobrick, Lee Sam Finn, Vahe G.\ Gurzadyan,
Nathan Johnson-McDaniel, Roger Penrose, Carlo Rovelli, David Sloan and
James Zibin for helpful discussions.
This work is supported in part by the NSF grant PHY0854743, Le Fonds qu\'eb\'ecois
de la recherche sur la nature et les technologies, the George A.\ and Margaret M.\
Downsbrough Endowment and the Eberly research funds of Penn State.

%\section*{References}

\end{document}